\documentclass[journal,12pt,onecolumn,draftclsnofoot]{IEEEtran}

\ifCLASSINFOpdf

\else

\fi
\usepackage[colorinlistoftodos]{todonotes}
\usepackage{amssymb}
\usepackage{amsmath,epsfig,amssymb,verbatim}
\usepackage{cite}
\usepackage[caption=false]{subfig}
\usepackage{array,algorithm,algorithmic}
\usepackage{comment}
\def\BibTeX{{\rm B\kern-.05em{\sc i\kern-.025em b}\kern-.08em
    T\kern-.1667em\lower.7ex\hbox{E}\kern-.125emX}}

\usepackage{tablefootnote}

\newtheorem{remark}{Remark}
\usepackage{enumerate}

\begin{document}

\title{Novel Spectrum Allocation Among Multiple Transmission Windows for Terahertz Communication Systems}

\author{Akram~Shafie,~\IEEEmembership{Graduate~Student~Member,~IEEE,} Nan~Yang,~\IEEEmembership{Senior~Member,~IEEE,} Chong~Han,~\IEEEmembership{Member,~IEEE,} and~Josep M. Jornet,~\IEEEmembership{Senior~Member,~IEEE}
\thanks{Akram Shafie and Nan Yang are with the School of Engineering, The Australian National University, Canberra, ACT 2601, Australia (e-mail: akram.shafie@anu.edu.au, nan.yang@anu.edu.au).}
\thanks{Chong Han is with the UM-SJTU Joint Institute, Shanghai Jiao Tong University, Shanghai 200240, China (e-mail: chong.han@sjtu.edu.cn). }
\thanks{Josep~M.~Jornet is with the Department of Electrical and Computer Engineering, Northeastern University, Boston, MA 02120, USA (e-mail: jmjornet@northeastern.edu). }
 }
\maketitle

\vspace{-15mm}
\begin{abstract}
This paper presents a novel spectrum allocation strategy for multiuser terahertz (THz) band communication systems when the to-be-allocated spectrum is composed of multiple transmission windows (TWs). This strategy explores the benefits of (i) allowing users to occupy sub-bands with unequal bandwidths and (ii) optimally avoiding using some spectra that exist at the edges of TWs where molecular absorption loss is high. To maximize the aggregated multiuser data rate, we formulate an optimization problem, with the primary focus on spectrum allocation. We then apply transformations and modifications to make the problem computationally tractable, and develop an iterative algorithm based on successive convex approximation to determine the optimal sub-band bandwidth and the unused spectra at the edges of TWs. Using numerical results, we show that a significantly higher data rate can be achieved by changing the sub-band bandwidth, as compared to equal sub-band bandwidth. We also show that a further data rate gain can be obtained by optimally determining the unused spectra at the edges of TWs, as compared to  avoiding using pre-defined spectra at the edges of TWs.
\end{abstract}

\begin{IEEEkeywords}
Terahertz communication, spectrum allocation, adaptive bandwidth, transmission windows.
\end{IEEEkeywords}

\section{Introduction}

The terahertz (THz) band communication (THzCom), operating in the electromagnetic spectrum ranging from 0.1 to 10 THz, is envisioned as an enabling technology to support emerging wireless applications that require an explosive amount of data in the sixth-generation (6G) and beyond systems. Specifically,  the enormous available bandwidths in the order of a few tens to a hundred gigahertz (GHz) and the extremely short wavelengths offer enormous potentials for wireless transmission~\cite{2021_Ching_IEEEComMagTHzfor2030}.
Despite the promise, THzCom encounters unique challenges that have never been addressed at lower frequencies, e.g., higher channel sparsity, severe spreading loss, and frequency-dependent molecular absorption loss~\cite{2020_WCM_THzMag_TerahertzNetworks}.
{These challenges need} to be wisely tackled for developing ready-to-use THzCom systems.

To fully harness the potentials of the THz band, the huge available bandwidth that spans across multiple ultra-wide transmission windows (TWs) needs to be optimally useds~\cite{20206GAkyidliz,HBM2}.
This necessitates the exploration of novel and efficient spectrum allocation schemes for THzCom system when the to-be-allocated spectrum of interest is composed of multiple TWs.
{In this work, we consider the entire spectrum regions that exist between neighboring molecular absorption coefficient peaks at the THz band as TWs~\cite{HBM2}}.
Multi-band-based spectrum allocation scheme has recently been studied for micro- and macro-scale THzCom systems with multiple TWs in the to-be-allocated spectrum of interest.
In this scheme, the spectrum of interest is divided into a set of non-overlapping sub-bands  which have a relatively small bandwidth, and then the sub-bands are utilized to serve the users in the system.
This scheme has the ability to efficiently allocate spectral resources while ensuring intra-band interference-free transmission.
{Triggered by this ability, multi-band-based spectrum allocation has been widely explored for micro- and macro-scale multiuser THzCom systems~\cite{HBM2,2020_Chong_InfoCom_DABM,2019_Chong_DABM2,2020ICC_NOMAforTHz,2021_THz_NOMA}.}

A novel sub-band assignment strategy for multi-band-based spectrum allocation was proposed  in~\cite{HBM2} to improve the throughput fairness among users in multiuser THzCom systems.
Moreover, spectrum allocation problems in the THz band backhaul network and non-orthogonal multiple access (NOMA) assisted THzCom systems were addressed in~\cite{2020_Chong_InfoCom_DABM}, and~\cite{2019_Chong_DABM2}, respectively.
{Furthermore,  efficient transmit power allocation and sub-band assignment algorithms were developed in \cite{2020ICC_NOMAforTHz} and \cite{2021_THz_NOMA}, respectively, for multiuser THzCom systems.
It is noted that the studies~\cite{HBM2,2020_Chong_InfoCom_DABM,2019_Chong_DABM2,2020ICC_NOMAforTHz,2021_THz_NOMA} focused on  multi-band-based spectrum allocation with equal sub-band bandwidth (ESB), where the spectrum of interest is divided into sub-bands with equal bandwidth.}
However,  since the molecular absorption loss varies considerably within the THz band, the variation among the sub-bands would be very high when ESB is considered. This variation can be reduced by exploring spectrum allocation with adaptive sub-band bandwidth (ASB), which allows to change the sub-band bandwidth. This can eventually lead to an overall improvement in the data rate performance.

Motivated by the potentials of ASB, the multi-band-based spectrum allocation with ASB was first introduced in our previous study~\cite{akram2021TCOM}. However, the design in~\cite{akram2021TCOM} is only applicable when the spectrum of interest exists entirely within a specific region of a TW. When the to-be-allocated spectrum of interest spans across multiple TWs,  on one hand, the number of sub-bands that exist within each TW needs to be optimally decided. On the other hand, the set of users that are served by each TW and the sub-band assignment policy need to be identified.
These challenges have not been touched by~\cite{akram2021TCOM}, but need to be carefully addressed to exploit the potentials of the huge available bandwidth that spans across multiple TWs at the THz band. This motivates this work.

We note that since molecular absorption loss is high at the edges of TWs, it may be beneficial if some spectra at the edges of TWs are not used during spectrum allocation.
Despite so, previous studies including ours~\cite{akram2021TCOM,HBM2,2020_Chong_InfoCom_DABM,2019_Chong_DABM2,2020ICC_NOMAforTHz,2021_THz_NOMA} considered that the entire spectrum existing within the to-be-allocated spectrum of interest is fully utilized. {Several recent studies \cite{2020_WCNC_HangNan,HBM1} considered to avoid some spectra at the edges of TWs. However, the designs in \cite{2020_WCNC_HangNan,HBM1} considered the unused spectra at the edges of TWs to be pre-defined, i.e., the unused spectra are arbitrarily chosen and fixed,  which may not guarantee the optimal performance.}

In this work, we design a novel multi-band-based spectrum allocation strategy when the to-be-allocated spectrum of interest is composed of multiple TWs. Specifically, we explore the benefits of ASB and optimally avoiding some spectra that exist at the edges of TWs during spectrum allocation.
With these considerations in mind and focusing on an indoor THzCom scenario where a single access point (AP) supports the downlink of multiple users, we formulate a spectrum allocation problem. To  analytically solve this problem, we propose transformations, as well as  modifications, and arrive at an approximate convex problem. Subsequently, we develop an iterative algorithm based on the successive convex approximation technique to solve the approximate convex problem.
Using numerical results, we show that when the to-be-allocated spectrum of interest is composed of multiple TWs, enabling ASB achieves a significantly higher data rate  (between $9\%$ and $15\%$) as compared to adopting ESB.
We also show that an additional gain of up to $5\%$ can be obtained by optimally determining the unused spectra at the edges of TWs, as compared to  avoiding using pre-defined spectra that exist at the edges of TWs, especially when the power budget constraint is stringent.
{We finally illustrate that the feasibility region of the spectrum allocation problem can be improved by considering ASB and optimally avoiding some spectra that exist at the edges of TWs.}

\section{System Model and Problem Formulation}

We consider a  three-dimensional (3D) indoor THzCom system where a single AP, which is located at the center of the ceiling of the indoor environment, supports the downlink of users which demand high data rates.
We consider that the users are stationary and distributed uniformly on the floor in this environment.
We denote $\mathcal{I}$ as the set for the users and further denote $l_{i}$ and $d_{i}{=}\sqrt{h_{\epsilon}^2 {+}l_{i}^2}$ as the horizontal and 3D distances of the link between the AP and the $i$th user, respectively, where $h_{\epsilon}$ is the fixed difference between the heights of AP and the users~\cite{akram2020JSAC}.

\subsection{THz Spectrum}

\begin{figure}[!t]
\centering
\includegraphics[width=0.8\columnwidth]{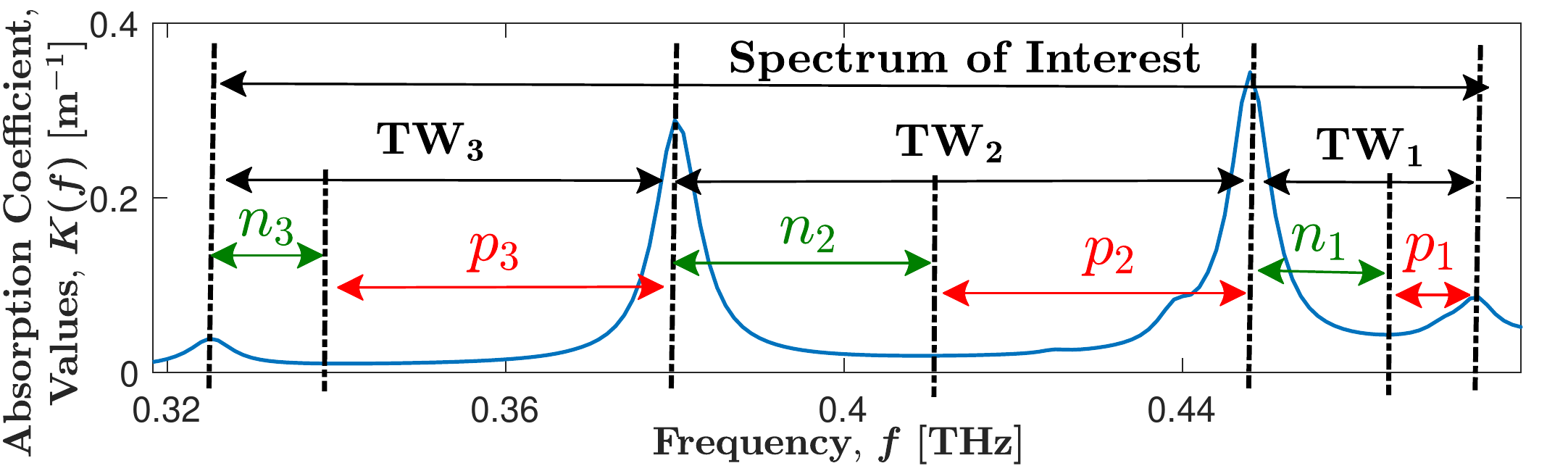}
\vspace{-4mm}
\caption{Illustration of TWs, PACSRs,  and NACSRs that exist between 0.32 THz and 0.48 THz, with $\{\mathrm{TW}_{1}, \mathrm{TW}_{2}, \mathrm{TW}_{3}\}$, $\{p_1,p_2,p_3\}$, and  $\{n_1,n_2,n_3\}$ denoting the TWs, PACSRs, and NACSRs, respectively. }\label{Fig:THzBand1} 
\end{figure}

The intermittent molecular absorption loss peaks that are observed throughout the THz band at different frequencies divides the entire THz band into multiple ultra-wideband transmission windows (TWs)~\cite{akram2021TCOM}. {In this work, the entire spectrum regions that exist between neighboring molecular absorption coefficient peaks are denoted by TWs~\cite{HBM2}.}
We focus in this work on the scenario where the to-be-allocated spectrum of interest is composed of multiple TWs, as shown in Fig. \ref{Fig:THzBand1}. For analytical purposes, we consider that each TW is composed of multiple positive absorption coefficient slope regions (PACSRs) and negative absorption coefficient slope regions (NACSRs). Here, PACSRs and NACSRs are defined as the regions within  TWs with increasing and decreasing molecular absorption coefficients, respectively. Considering this, we find that the to-be-allocated spectrum of interest is made of multiple PACSRs and NACSRs, as shown in Fig. \ref{Fig:THzBand1}.
We denote $\mathcal{R}_\textrm{P}{=}\{p_1,p_2, \cdots  \}$ and  $\mathcal{R}_\textrm{N}{=}\{n_1,n_2, \cdots  \}$  as the sets of the PACSRs and the NACSRs in the spectrum of interest, respectively. We further denote $\mathcal{R}{=}\{r, r\!\in\! \mathcal{R}_\textrm{P}\cup \mathcal{R}_\textrm{N} \}$ as the set of all regions in the spectrum of interest.

\begin{figure}[!t]
\centering
\includegraphics[width=0.8\columnwidth]{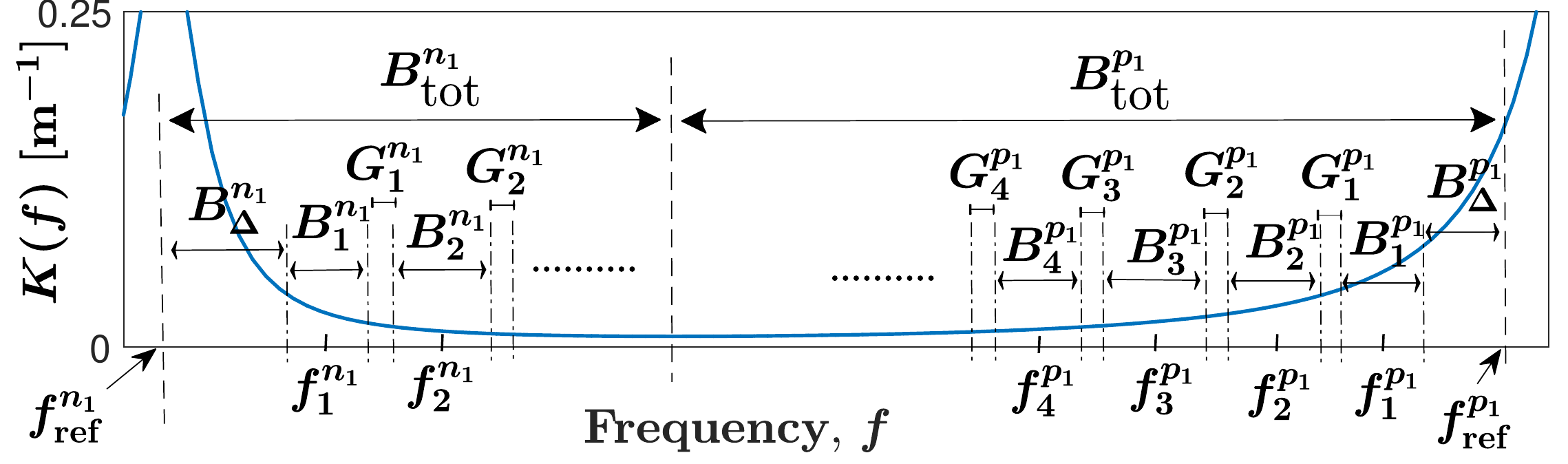}
\vspace{-4mm}
\caption{Illustration of sub-band arrangement within a PACSR and an NACSR.}\label{Fig:THzBand2} 
\end{figure}

We focus on multi-band-based spectrum allocation. Thus, each region within the spectrum of interest is divided into sub-bands that are separated by guard bands. 
An illustration of the arrangement of sub-bands within a PACSR and an NACSR is depicted in Fig. \ref{Fig:THzBand2}.
We denote  $\mathcal{S}^r{=}\{1,2, \cdots, s, \cdots \}$ as the set of the sub-bands that exist in the region $r$ and further denote $B_{s}^{r}$ and $f_{s}^{r}$  as the bandwidth and the center frequency of the $s$th sub-band that exist in the region $r$, respectively.
For notational convenience, we consider that sub-bands are labeled such that $f_1^{r}> f_2^{r}> \cdots > f_{s}^{r}> \cdots$ when $r\!\in\! \mathcal{R}_\textrm{P}$ and $f_1^{r}\!<\! f_2^{r}\!<\! \cdots \!<\! f_{s}^{r}\!<\! \cdots$ when $r\!\in\! \mathcal{R}_\textrm{N}$.
Thus, $f_{s}^{r}$ can be expressed as 
\begin{align}\label{Equ:f_s}
  f_{s}^{r}&=f_{\textrm{ref}}^{r} - \eta_r\left(B^r_{\Delta}+\sum\nolimits_{k=1}^{s-1}B_{k}^{r}+0.5 B_{s}^{r}+\sum\nolimits_{k=1}^{s-1}G_{k}^{r}\right)\notag\\
  &=f_{\textrm{ref}}^{r} - \eta_r\left(B^r_{\Delta}+\sum\nolimits_{k\in \mathcal {S}}\left(a_{1,s,k}B_{k}^{r}+a_{2,s,k}G_{k}^{r}\right)\right),
\end{align}
where $s\!\in\! \mathcal{S}^r,r\!\in\! \mathcal {R}$, $f_{\textrm{ref}}^{r}$  is the end-frequency of the region $r$ when $r\!\in\! \mathcal{R}_\textrm{P}$ and the start-frequency of the region $r$ when $r\!\in\! \mathcal{R}_\textrm{N}$,  as shown in Fig. \ref{Fig:THzBand2}. In addition,   $\eta_r{=}1$ when $r\!\in\!~\mathcal{R}_\textrm{P}, \eta_r{=}-1$ when $r\!\in\!~\mathcal{R}_\textrm{N} $, $B^r_{\Delta}$ is the bandwidth of the unused spectrum that exists in the region~$r$, and
$G_{s}^{r}$ is the bandwidth of the guard band that exists between the $s$th and the $(s-1)$th the sub-band in the region $r$. Moreover, $a_{1,s,k}{=}1,~\textrm{if}~k<s$, $a_{1,s,k}{=}0.5,~\textrm{if}~k{=}s$, $a_{1,s,k}{=}0,~\textrm{otherwise}$~$\forall s,k \!\in\! \mathbb{Z}^+$ and
$a_{2,s,k}{=}1, ,~\textrm{if}~ k < s$, $a_{2,s,k}{=}0,~\textrm{otherwise}$~$\forall s,k \!\in\! \mathbb{Z}^+$.
{In this work, we optimally determine the value of $B_{s}^{r}$ by solving the optimization problem that will be formulated in the subsequent section.}
{Also, we consider that the guard band bandwidth is fixed and is equal to $B_{\textrm{g}}$, i.e., $G_{s}^{r}{=}B_{\textrm{g}}, ~ \forall s\!\in\! \mathcal{S}^r, r\!\in\! \mathcal {R}$}{\footnote{{
To further improve the performance of the spectrum allocation strategy, it may be beneficial to optimally design the values of $G_{s}^{r}$, which will be considered in our future work.}}}.

 \begin{remark} \label{Rem:1}
As shown in Figs. \ref{Fig:THzBand1} and \ref{Fig:THzBand2}, the values of the molecular absorption coefficient are very high  at the edges of TWs. Due to this, it may be beneficial if some spectrum that exists at the edges of TWs are not utilized during spectrum allocation. To optimally determine the unused spectra at the edges of TW, we introduce a design variable $B^r_{\Delta}$ in our spectrum allocation strategy, where  $B^r_{\Delta}$ denotes the unused spectra at the edge of each region where the molecular absorption coefficient is very high.
It is reflect in~\eqref{Equ:f_s} and depicted in Fig. \ref{Fig:THzBand2}.
\end{remark}

With all the aforementioned considerations, we express the total bandwidth available within the region $r$ as
 \begin{align}\label{OptProbOrg-G} 
 B_{\textrm{tot}}^{r}=B^r_{\Delta}+\sum\nolimits_{s\in \mathcal{S}^r}B_{s}^{r}+\sum\nolimits_{s\in \mathcal{S}^r}G_{s}^{r},~~~ \forall r \in \mathcal{R}. \end{align}

We assume that the users in the system are served by separate sub-bands~\cite{HBM2,akram2021TCOM}. This assumption ensures intra-band interference-free data transmission and eliminate the hardware complexity and signal processing overhead caused by frequency reuse in the system{\footnote{{A higher spectral efficiency can be achieved by allowing (i) users to occupy multiple sub-bands and/or (ii)  sub-band reuse during spectrum allocation, which will be considered in our future work.}}}. Under this assumption, we set the total number of sub-bands in the spectrum of interest to be equal to the number of users in the system, i.e.,
\begin{align}\label{OptProbOrg-I3}
   & \sum\nolimits_{r \in \mathcal{R}} |\mathcal{S}^r|=|\mathcal{I}|.
\end{align}

\noindent Although the total number of sub-bands in the spectrum of interest, i.e., $\sum\nolimits_{r \in \mathcal{R}} |\mathcal{S}^r|$, is defined by \eqref{OptProbOrg-I3}, the number of sub-bands that exist within each region, i.e.,  $|\mathcal{S}^r|,~  \forall r \!\in\! \mathcal{R}$, is unknown. Thus, it should be carefully decided
such that the spectra available within the regions are utilized optimally,
while ensuring that the constraint \eqref{OptProbOrg-I3} is satisfied.

Let us denote $x^r_{i,s}$ as the sub-band assignment indicator variable such that $x^r_{i,s}{=}1$ if~the $s$th sub-band in~the region $r$ is~assigned to the $i$th user and $x^r_{i,s}{=}0$, otherwise, $ \forall i\!\in\! \mathcal {I}, s\!\in\! \mathcal{S}^r, r\!\in\! \mathcal {R}$. Under the assumption that the users in the system are served by separate sub-bands, we have
\begin{subequations}\label{OptProbOrg} 
\begin{alignat}{2}
 \sum\nolimits_{r\in \mathcal {R}} \sum\nolimits_{s\in \mathcal{S}^r}x^r_{i,s}  &=1, ~~~\forall i\in \mathcal {I}, \label{OptProbOrg-I1}\\
    \sum\nolimits_{i\in \mathcal {I}}x^r_{i,s}  &= 1, ~~~ \forall s\in \mathcal{S}^r,r\in \mathcal {R}, \label{OptProbOrg-I2}
\end{alignat}
\end{subequations}
where  \eqref{OptProbOrg-I1} ensures  that each user in the system is assigned with one sub-band and \eqref{OptProbOrg-I2} ensures that each sub-band in the spectrum of interest is assigned to one user.

\subsection{Achievable Data Rate}

\label{Sec:Channel}

{Considering the spreading loss and the frequency-selective molecular absorption loss at the THz band, the data rate achieved by the $i$th user in  the $s$th sub-band in region $r$ is obtained as an integral of capacity on frequency, given by~\cite{HBM2}
\begin{align} \label{Equ:Rateijs1} 
&R_{i,s}^r=\!\!\!\!\!\!\int\limits_{f^r_s-\frac{1}{2}B_{s}^r}^{f_s+\frac{1}{2}B_{s}^r}   \!\!\!\!\!\!C(f)~\textrm{d}f=\!\!\!\!\!\!\!\!\int\limits_{f^r_s-\frac{1}{2}B_{s}^r}^{f^r_s+\frac{1}{2}B_{s}^r}   \!\!\!\!\!\!\!\log_{2}\left(\!1{+}\frac{P_{i} g_{\textrm{A}}g_{\textrm{U}} c^2 e^{-K(f) d_i}}{(4 \pi f d_i)^2\Omega_s^r}\!\right)\textrm{d}f,
\end{align}
where $i\!\in\! \mathcal {I}, s\!\in \!\mathcal {S}^r \!,\! r\!\in\! \mathcal {R}$, $C(f)$ is the capacity when the frequency is $f$,} $P_{i}$ is the transmit power allocated to the $i$th user, $g_{\textrm{A}}$ and $g_{\textrm{U}}$ are the antenna gains at the AP and users, respectively, $c$ is the speed of light, and $K(f)$ is the molecular absorption coefficient at $f$\footnote{The impact of inter-band interference (IBI) is not considered in this work due the fact that there exist  IBI suppression schemes that can suppress IBI with minimal throughput degradation and waveform designs that can minimize power leakages to adjacent bands~\cite{akram2021TCOM,2020_WCNC_HangNan}.}. {Following~\cite{2021_JSAC_VariableBandwidth,akram2021TCOM}, the noise power is calculated as a function of the sub-band bandwidth. Specifically, noise power of the $s$th sub-band in region $r$, $\Omega_s^r$, is obtained as $\Omega_s^r=N_0 B_{s}^r$, where $N_{0}$ is the noise power density~\cite{2021_JSAC_VariableBandwidth,akram2021TCOM}.} We assume that users in the system are equipped with tunable front-ends such that they can dynamically adjust their operating frequency and bandwidth to match those of their assigned sub-bands.

We note that although the values of  $K(f)$ can be found by using the HITRAN database~\cite{2008_Hitran}, {line by line radiative transfer model (LBLRTM)~\cite{Kfcalculation1_2020LBLRTM},} or the International Telecommunication
Union (ITU) model~\cite{2019ITU}, there does not exist a tractable expression  that maps $f$ to $K(f)$ within the entire THz band and can be used to analytically model spectrum allocation problems with ASB~\cite{akram2021TCOM}. With this in mind, through curve fitting,
we model the values of  $K(f)$ corresponding to each region using separate exponential functions of $f$. Specifically, we approximate $K(f)$ in the region $r$ as
\begin{equation}\label{Equ:CurveFit} 
  \hat{K^r}(f)=e^{\sigma^r_{1} +\sigma^r_{2} f}+\sigma^r_{3},
\end{equation}
where  $\min\{f_{\textrm{ref}}^{r}, f_{\textrm{ref}}^{r} -\eta_r B_{\textrm{tot}}^r\}\leqslant f \leqslant \max\{f_{\textrm{ref}}^{r}, f_{\textrm{ref}}^{r} -\eta_r B_{\textrm{tot}}^r\}$, $r\!\in\! \mathcal {R}$
and  $\{\sigma^r_{1},\sigma^r_{2},\sigma^r_{3}\}$ are the model parameters obtained for the region $r$. It is noted that $\sigma^r_{2} >0~\textrm{when}~r\!\in\! \mathcal{R}_{\textrm{P}}$ and $\sigma^r_{2} <0~\textrm{when}~r\!\in\! \mathcal{R}_{\textrm{N}}$.
Thereafter, by substituting \eqref{Equ:CurveFit} into \eqref{Equ:Rateijs1}, $R_{i,s}^r$ is obtained as  
{
\begin{align} \label{Equ:Rateijs2} 
R_{i,s}^r&=\!\!\int\limits_{f^r_s-\frac{1}{2}B_{s}^r}^{f^r_s+\frac{1}{2}B_{s}^r}   \!\!\!\!\!  \log_{2}\left(\!1{+}\frac{P_{i} \varrho e^{-\left(e^{\sigma^r_{1} +\sigma^r_{2} f}+\sigma^r_{3}\right) d_i}}{(f d_i)^2B_{s}^r}\!\right) \textrm{d}f,
\end{align}}
where $\varrho {=} \frac{g_{\textrm{A}}g_{\textrm{U}}}{N_{0}}\left(\frac{c}{4 \pi }\right)^{2}$ and $f_s^r$ is given in \eqref{Equ:f_s}.
{When $K(f)$ behaves according to~\eqref{Equ:CurveFit}, $C(f)$ within the PACSR (or the NACSR) decreases (or increases) monotonically with frequency. We approximate this monotonically decreasing (or increasing) behaviour of $C(f)$ using a piecewise linear function. This approximation allows to modify $R_{i,s}^r$ in~\eqref{Equ:Rateijs2} into a simplified expression, given by
\begin{align} \label{Equ:Rateijs3} 
\!\!R_{i,s}^r&{=}B_{s}^r  \log_{2}\left(\!1{+}\frac{P_{i} \varrho e^{-\left(e^{\sigma^r_{1} +\sigma^r_{2} f_s^r}+\sigma^r_{3}\right) d_i}}{(f_s^r d_i)^2B_{s}^r}\!\right).
\end{align}}

\noindent
Finally, the achievable data rate of the $i$th user is obtained as
\begin{align} \label{Equ:Rate1} 
R_{i}&= \sum\nolimits_{r\in \mathcal {R}} \sum\nolimits_{s\in \mathcal{S}^r}x^r_{i,s} R^r_{i,s},
~~~ \forall i\in \mathcal {I}.
\end{align}

\section{Optimal Spectrum Allocation}

In this section, we present the  multi-band-based spectrum allocation problem with ASB when the to-be-allocated spectrum of interest is composed of multiple TWs.
We aim to maximize the sum data rate among all users under given sub-band bandwidth, power, and rate constraints.
Mathematically, this problem is formulated as
\begin{subequations}\label{OptProbOrg} 
\begin{alignat}{2}
& \mathbf {P^{o}:} \quad  \underset{\substack{ x^r_{i,s},P_{i},B^r_{s},\\ B^r_{\Delta}, \forall i ,s, r}}{\textrm{maximize}}
 \quad \sum_{i\in \mathcal {I}}R_{i}                                             \label{OptProbOrg-A} \\
&\quad  \quad  \quad      \textup{subject to}   \quad
    \eqref{OptProbOrg-G}, \eqref{OptProbOrg-I3}, \eqref{OptProbOrg-I1}, \eqref{OptProbOrg-I2}, \notag \\
&\quad \quad\quad  \quad  \quad  \quad \quad  \quad   \sum\nolimits_{i\in \mathcal {I}} P_{i} \leqslant P_{\textrm{tot}}, \label{OptProbOrg-B} \\
&\quad \quad \quad  \quad  \quad  \quad \quad  \quad 0 \leqslant P_{i}\leqslant P^{\textrm{max}}, ~~ \forall i\in \mathcal{I},                                  \label{OptProbOrg-B2} \\
&\quad \quad \quad  \quad \quad \quad \quad \quad R_{i} \geqslant  R_{\textrm{thr}}, ~~~ \forall i\in \mathcal {I},                           \label{OptProbOrg-D}\\
&\quad \quad \quad \quad \quad \quad \quad \quad 0\leqslant B_{s}^{r} \leqslant B_{\textrm{max}},  ~~~ \forall s\in \mathcal {S}^r, r \in \mathcal{R},                       \label{OptProbOrg-E}\\
&\quad \quad \quad \quad \quad \quad\quad \quad 0\leqslant B^r_{\Delta} \leqslant B_{\textrm{tot}}^{r},  ~~~ \forall  r \in \mathcal{R} ,                       \label{OptProbOrg-E2}\\
&\quad \quad \quad \quad \quad \quad \quad \quad  x^r_{i,s}
 \in \{0,1\} ,~~~  \forall i\in \mathcal {I},s\in \mathcal{S}^r, r\in \mathcal {R}.      \label{OptProbOrg-J}
\end{alignat}
\end{subequations}

\noindent The problem consists of sub-band assignment, sub-band bandwidth allocation, determination of the unused spectra at the edges of TW, and power control. The constraint \eqref{OptProbOrg-B} reflects the power budget at the AP,  \eqref{OptProbOrg-B2} ensures that the power allocated to each user is upper bounded by  $P^{\textrm{max}}$, \eqref{OptProbOrg-D} ensures that the data rate achieved by each user is greater than the threshold~$R_{\textrm{thr}}$, and \eqref{OptProbOrg-E} ensures that the bandwidth of each sub-band is upper bounded by $B_{\textrm{max}}$.

\subsection{Solution Approach}

We now present the solution to the  optimization problem $\mathbf {P^{o}}$.
We note that it is extremely difficult to use traditional optimization theory techniques to analytically solve $\mathbf {P^{o}}$ in its current form. This is because the number of sub-bands that exist within each region, i.e.,  $|\mathcal{S}^r|,~ \forall r \!\in\! \mathcal{R}$, is unknown. This leads to the fact that the number of design variables $x^r_{i,s}$ and $B^r_{s}$  are unknown. Therefore, we need to render the problem $\mathbf {P^{o}}$ computationally tractable. Considering this, we introduce  modifications to the problem $\mathbf {P^{o}}$,  as follows:

We first consider that there exist $|\mathcal{I}|$ sub-bands within each region, i.e., $|\mathcal{S}^r|{=}|\mathcal{I}|$. %
This implies that there would be $|\mathcal{R}|\times|\mathcal{I}|$ sub-bands in the entire spectrum of interest. {Out of these $|\mathcal{R}|\times|\mathcal{I}|$ sub-bands, we ensure that only some of the  sub-bands are used by at most one user, while the remaining sub-bands are not assigned to any user.}
To reflect this modification, we replace \eqref{OptProbOrg-I3} and \eqref{OptProbOrg-I2} in $ \mathbf {P^{o}}$ with
\begin{subequations} 
\begin{alignat}{2}
&{|\mathcal{S}^r|{=}|\mathcal{I}|,  ~~~~~~~~~~~\forall r\in \mathcal {R},}                                                                                              \label{OptProbOrg-I2X}\\
&\sum\nolimits_{i\in \mathcal {I}}x^r_{i,s}  \leqslant 1, ~~~\forall s\in \mathcal{S}^r,r\in \mathcal {R}.                                                                                                  \label{OptProbOrg-I2X2}
\end{alignat}
\end{subequations}

 After this modification, although the intractability of $\mathbf {P^{o}}$ caused by the unknown number of design variables corresponding to each region can be solved, a new challenge arises due to the existence of guard bands in each region. Specifically, our spectrum allocation strategy considers that guard bands with bandwidth of $B_{\textrm{g}}$ exist between sub-bands. Thus, $(|\mathcal{I}|{-}1)$ guard bands appear in each region as each region consists $|\mathcal{I}|$ sub-bands after the modification. This would result in guard bands existing even between sub-bands which have zero bandwidth, which must to be avoided.
To overcome this challenge, we first consider that the bandwidths of guard bands, $G_s^r,~\forall s\!\in\! \mathcal{S}^r,r\!\in\! \mathcal {R}$, are variables such that $0\leqslant G_s^r\leqslant B_{\textrm{g}}$.
Then, we ensure that there exists exactly one guard band with the bandwidth of $B_{\textrm{g}}$ between sub-bands that have non-zero bandwidth, and the remaining guard bands have zero bandwidth. This is achieved by ensuring that the guard band between the $s$th and the $(s{-}1)$th sub-band in a given region has non-zero bandwidth only if the $s$th sub-band of this region has non-zero bandwidth.
Mathematically, this is expressed as
\begin{align}\label{Equ:Bg1} 
G_s^r=\begin{cases}
B_{\textrm{g}}, & B_s^r\neq 0 ,\\
0, & \textrm{otherwise},
\end{cases} ~~\forall s\in \mathcal{S}^r,r\in \mathcal {R}.
\end{align}
We then note that $B_s^r\neq 0$ when $\sum\nolimits_{i\in \mathcal {I}}x^r_{i,s}{=}1$. Considering this,  \eqref{Equ:Bg1} can be equivalently  expressed as
\begin{align}
\!\!\! B_{\textrm{g}}\sum\nolimits_{i\in \mathcal {I}}x^r_{i,s} \leqslant G_{s}^{r} \leqslant B_{\textrm{g}} \sum\nolimits_{i\in \mathcal {I}}x^r_{i,s} , ~~~ \forall s\in \mathcal{S}^r,r\in \mathcal {R}.                                                                                                  \label{Equ:Bg2}
\end{align}

\noindent With these aforementioned modifications,
we transform $\mathbf {P^{o}}$  into the following tractable optimization problem,  given by
\begin{alignat}{2}
& \mathbf {\bar{P}^{o}:}\quad  \underset{\substack{x^r_{i,s},P_{i}, B^r_{\Delta},\\   B^r_{s}, G_{s}^{r}, \forall i ,s, r}}{\textrm{minimize}}
& & -\sum_{i\in \mathcal {I}}R_{i}                                               \label{OptProbOrg-A} \\
&\quad  \quad  \quad      \textup{subject to}
 & & \quad \eqref{OptProbOrg-G}, \eqref{OptProbOrg-I1},   \eqref{OptProbOrg-B}- \eqref{OptProbOrg-J}, \eqref{OptProbOrg-I2X}, \eqref{OptProbOrg-I2X2}. \eqref{Equ:Bg2}. \notag
\end{alignat}
\noindent

Next, we note that although  $\mathbf {\bar{P}^{o}}$ is tractable, it is non-convex in its current form. {In particular, the non-convexity in $\mathbf {\bar{P}^{o}}$ arises due to (i) $R_{i}$ in \eqref{Equ:Rate1} is non-concave  w.r.t. the design variables $B_{\nu}^{\tau}$, $B^{\tau}_{\Delta}$,$G_{\nu}^{\tau}$, $\forall~\!\nu\!\!\in\!\!\mathcal{S}^r\!, \tau\!\!\in\!\! \mathcal{R}$,} and (ii) the  design variable $x^r_{i,s}$ is binary.
{Considering this, first, to handle the non-concavity of $R_{i}$,
we introduce the following substitutions for $B^r_{s}$, $B^r_{\Delta}$, $G^r_{s}$, given~by}
\begin{subequations}\label{OptProbOrg}
\begin{alignat}{2}
 \label{Equ:Trans1}
  B^r_{s}&=\xi^r_{1,s}+\omega^r_{1,s}\log(\varsigma^r_{1,s} Z^r_{1,s}), ~~\forall s\in \mathcal{S}^r, r \in \mathcal{R}, \\
    \label{Equ:Trans2}
  B^r_{\Delta}&=\xi^r_{1,0}+\omega^r_{1,0}\log(\varsigma^r_{1,0} Z^r_{1,0}), ~~\forall r \in \mathcal{R},\\
  \label{Equ:Trans3}
  G^r_{s}&=\xi^r_{2,s}+\omega^r_{2,s}\log(\varsigma^r_{2,s} Z^r_{2,s}), ~~\forall s\in \mathcal {S}^r, r \in \mathcal{R},
\end{alignat}
\end{subequations}

\noindent
where {$\xi^r_{\phi,s},\omega^r_{\phi,s},\varsigma^r_{\phi,s}$ are real constants with $\phi\!\in\!\Phi{=}\{1,2\}$} and $Z^r_{\phi,s}$, $\forall \phi\!\in\!\Phi,~{s}\!\in\! \mathcal{S}_{\phi}^r, r \!\in\! \mathcal{R}$ are the new design variables that replace $B^r_{s}$, $B^r_{\Delta}$, $G^r_{s}$  in the optimization problem, with $\mathcal {S}^r_{1}{=}\mathcal {S}^r \cup \{0\}$ and
$\mathcal {S}^r_{2}{=}\mathcal {S}^r$.
Considering these replacements, we transform  $\mathbf {\bar{P}^{o}}$ into the following equivalent problem, given by
\begin{subequations}\label{OptProbOrg3}
\begin{alignat}{2}
& \mathbf {\tilde{P}^{o}:} \underset{\substack{x^r_{i,s},P_{i}, Z^r_{\phi,s},\\  \forall i ,s, r,\phi}}{\textrm{minimize}}
 \quad -\sum_{i\in \mathcal {I}}\sum_{r\in \mathcal {R}} \sum_{s\in \mathcal{S}^r}x^r_{i,s} R^r_{i,s}                                             \label{OptProbOrg-A} \\
&  \quad \quad\textup{subject to}
    \quad   \eqref{OptProbOrg-I1}, \eqref{OptProbOrg-B}- \eqref{OptProbOrg-D}, \eqref{OptProbOrg-I2X}, \eqref{OptProbOrg-J}, \eqref{OptProbOrg-I2X2} ,  \notag \\
 &\quad\quad  \!\!\prod_{\phi\in\Phi}\!\!\! \prod_{~s\in \mathcal {S}_{\phi}^r\!\!\!}\!\!(Z_{\phi,s}^r)^{\omega^r_{\phi,s}}\!{-}\!\!\! \prod_{\phi\in\Phi} \!\!\! \prod_{~s\in \mathcal {S}_{\phi}^r\!\!\!} \!\! \frac{e^{\!\!B^r_{\textrm{tot}}\!{-}\!\!\!\sum\limits_{\phi\in \Phi} \sum\limits_{s\in \mathcal {S}^r} \!\xi_{\phi,s}^r}}{(\varsigma_{\phi,s}^r)^{\omega^r_{\phi,s}}}\!\!\leqslant\! 0,    ~\forall r \!\in\! \mathcal{R},  \label{OptProb2Sol2-C}\\
&  \quad\quad   Z^r_{\textrm{min},\phi,s} \leqslant Z^r_{\phi,s} \leqslant Z^r_{\textrm{max},\phi,s},\forall s\in \mathcal {S}^r, r\in \mathcal {R}, \phi\in \Phi, \label{OptProb2Sol2-A}
\end{alignat}
\end{subequations}
where $Z^r_{\textrm{min},1,s}{=}Z^r_{\textrm{ref},1,s}e^{\frac{\delta}{\omega^r_{1,s}}}$\!, $\delta$ is a very small positive number,
$Z^r_{\textrm{ref},\phi,s}{=}(\varsigma^r_{\phi,s})^{-1}e^{\frac{-\xi^r_{\phi,s}}{\omega^r_{\phi,s}}}$\!\!,
$Z^r_{\textrm{max},1,s}{=}Z^r_{\textrm{ref},1,s}e^{\frac{B_{\textrm{max}}}{\omega^r_{1,s}}}$\!, and
$Z^r_{\textrm{min},2,s}{=}$ $Z^r_{\textrm{max},2,s}{=}~\!Z^r_{\textrm{ref},2,s}\!\big(1{+}\!\sum\limits_{i\in \mathcal {I}}x^r_{i,s}\big(e^{\frac{B_g}{\omega^r_{2,s}}}{-}1\!\big)\big)$.
In~$\mathbf {\tilde{P}^{o}}$,  $R_{i,s}^r$ is given by
\begin{align} \label{Equ:Rateijs3}
\!\!R_{i,s}^r&{=}B_{s}^r  \log_{2}\!\!\left(\!1{+}\frac{P_{i} \varrho e^{-\left(e^{\sigma^r_{1} +\sigma^r_{2} \hat{f}_s^r}+\sigma^r_{3}\right) d_i}}{(\hat{f}_s^r d_i)^2 \left(\xi^r_{1,s}{+}\omega^r_{1,s}\log(\varsigma^r_{1,s} Z^r_{1,s})\right)}\!\right) ,
\end{align}
with $\hat{f}_s^r{=}f_{\textrm{ref}}^{r} {-} \eta_r \!\! \sum\nolimits_{\phi\in \Phi} \sum\nolimits_{k\in \mathcal {S}^r_{\phi}}\!\!a_{\phi,s,k}\!\left(\xi^r_{\phi,k}{+}\omega^r_{\phi,k}\log(\varsigma^r_{\phi,k} Z^r_{\phi,k})\right)$. We also note that, in $\mathbf {\tilde{P}^{o}}$,  $\omega^r_{s}$ is selected such that $1/\omega^r_{s}>\bar{\omega}^r$, $\forall$ ${s}\!\in\! \mathcal{S}_{\phi}^r, r \!\in\! \mathcal{R}$, $\phi\!\in\!\Phi$, to ensure the convexity of the objective function in $\mathbf {\tilde{P}^{o}}$ and the constraint function \eqref{OptProbOrg-D} w.r.t. $Z^{\tau}_{\phi,\nu}$, since $R^r_{i,s}$ in \eqref{Equ:Rateijs3} is concave w.r.t. $Z^{\tau}_{\phi,\nu}$ only when $1/\omega^r_{\nu}>\bar{\omega}^r$~\cite{akram2021TCOM}. Here, $\bar{\omega}^r{=}|\sigma^r_{2}|\left(D\hat{K}^r(f^r_{\textrm{ref}})e^{D\sigma^r_3}-1\right)$, $D{>}d_{\textrm{max}}$, and $d_{\textrm{max}}$ is the maximum of the link distances, i.e., $d_{\textrm{max}}{=}\max_{i\in \mathcal{I}}\{ d_{i}\}$.
{We clarify that for any selected combination of real values for $\xi^r_{\phi,s},\omega^r_{\phi,s},\varsigma^r_{\phi,s}$, the feasible region of the term $\xi^r_{\phi,s}+\omega^r_{\phi,s}\log(\varsigma^r_{\phi,s} Z_{s})$ is $\mathbb{R}$, as long as $Z^r_{\phi,s}{\in}\mathbb{R}$.
Thus, the constraint introduced on $\omega^r_{\phi,s}$ does not add any restriction on the domain of the problem $\mathbf {\tilde{P}^{o}}$.}

We now handle the non-convexity of $\mathbf {\tilde{P}^{o}}$ caused by the binary constraint function \eqref{OptProbOrg-D}.
{To this end, we first relax the binary variables in $\mathbf {\tilde{P}^{o}}$ into real variables~\cite{BinaryTrans1,2014_PenaltyFactor2,akram2021TCOM}. Let us denote $\tilde{x}^r_{i,s}$ as the real variables that would replace $x^r_{i,s}$ in the optimization problem. We then rewrite the binary constraint function  \eqref{OptProbOrg-J} equivalently as 
\begin{subequations}
\begin{alignat}{2}
& 0 \leqslant \tilde{x}^r_{i,s} \leqslant 1 ,~~~~~~ \forall i\in \mathcal {I},s\in \mathcal{S}^r, r\in \mathcal {R},   \label{OptProb1Sol1-B}\\
& \sum_{i\in \mathcal {I}}\! \sum_{r\in \mathcal {R}}\! \sum_{s\in \mathcal{S}^r}  \Big( \tilde{x}^r_{i,s} \big (1 - \tilde{x}^r_{i,s}  \big) \Big) \leqslant 0. \label{Equ:TransBin12}
\end{alignat}
\end{subequations}
Considering that the constraint~\eqref{Equ:TransBin12} is non-convex w.r.t $\tilde{x}^r_{i,s}$, we relax the constraint \eqref{Equ:TransBin12} and include it as a penalty function in  the objective function~\cite{BinaryTrans1,2014_PenaltyFactor2,akram2021TCOM}. In doing so, we transform $\mathbf {\tilde{P}^{o}}$ as
\begin{alignat}{2}\label{OptProbOrg4}
& \mathbf {\hat{P}^{o}:} \underset{\substack{\tilde{x}^r_{i,s},P_{i}, Z^r_{\phi,s},\\  \forall i ,s, r,\phi}}{\textrm{minimize}}
 \quad       \Psi^{(\kappa)}\left(\tilde{x}^r_{i,s},P_{i}, Z^r_{\phi,s}\right)                                      \\
& \quad\quad \textup{subject to}
    \quad  \tilde{ \eqref{OptProbOrg-I1}},  \eqref{OptProbOrg-B}- \eqref{OptProbOrg-D}, \tilde{\eqref{OptProbOrg-I2X}}, \tilde{\eqref{OptProbOrg-I2X2}}, \tilde{\eqref{OptProb2Sol2-C}}, \eqref{OptProb2Sol2-A}, \eqref{OptProb1Sol1-B} \notag,
\end{alignat}
where $\Psi^{(\kappa)}\left(\tilde{x}^r_{i,s},P_{i}, Z^r_{\phi,s}\right)=-\sum\limits_{i\in \mathcal {I}}\sum\limits_{r\in \mathcal {R}} \sum\limits_{s\in \mathcal{S}^r}\tilde{x}^r_{i,s} R^r_{i,s} +\Lambda F_{\textrm{p}}^{(\kappa)}$ is the objective function of $\mathbf {\hat{P}^{o}}$, $\Lambda \geqslant 0$ is the constant penalty factor, $F_{\textrm{p}}^{(\kappa)} {=}\sum_{i\in \mathcal {I}}\! \sum_{r\in \mathcal {R}}\! \sum_{s\in \mathcal{S}^r}  \!\!\Big(\!\big ((\tilde{x}^r_{i,s})^{(\kappa)}  \big)^{2} \!{+}\tilde{x}^r_{i,s} \big (1 {-} 2 (\tilde{x}^r_{i,s})^{(\kappa)}  \big)\! \Big)$, and $(\tilde{x}^r_{i,s})^{(\kappa)}$ is an approximate of $\tilde{x}^r_{i,s}$. Moreover, $\tilde{ \eqref{OptProbOrg-I1}}$, $\tilde{\eqref{OptProbOrg-I2X}}$, $\tilde{\eqref{OptProbOrg-I2X2}}$, and $\tilde{\eqref{OptProb2Sol2-C}}$ are the constraints obtained by replacing $x^r_{i,s}$ with $\tilde{x}^r_{i,s}$ in ${ \eqref{OptProbOrg-I1}}$, ${\eqref{OptProbOrg-I2X}}$,  ${\eqref{OptProbOrg-I2X2}}$, and \eqref{OptProb2Sol2-C}, respectively.}
We note that  $\mathbf {\hat{P}^{o}}$  is obtained as a standard convex optimization problem. 
Thus, by using the successive convex approximation technique, we propose an iterative algorithm to solve $\mathbf {\hat{P}^{o}}$, which is
summarized in Algorithm~\ref{Alg:Alg1}.

\begin{algorithm}[t]

\caption{Iterative approach for solving the problem $\mathbf {\hat{P}^{o}}$.}
\begin{algorithmic}[1] \label{Alg:Alg1}
\STATE $\!\!$\textbf{Initialization}: Set iteration count $\kappa{=}0$. Set initial point for $\tilde{x}^{r (\kappa)}_{i,s}{=}0.5, \, \forall i\in \mathcal {I}, r\in \mathcal {R}, s\!\in\! \mathcal {S}_r$.  Select a high penalty factor $\Lambda{=}200$ and a low tolerance value $\epsilon{=}10^{-6}$.
\WHILE {$F_{\textrm{p}}^{(\kappa)}  \geqslant \epsilon$}
   \STATE $\!\!\!\!\!\!\!\!\!$ Solve  $\mathbf {\hat{P}^{o}}$ in~\eqref{OptProbOrg4} using the point $\tilde{x}^{r (\kappa)}_{i,s}$,  and obtain $ P_{i}^{*}, \tilde{x}_{i,s}^{r*} ,Z_{\phi,s}^{r*} $ and $\Psi^{(\kappa)}\left(\tilde{x}^{r*}_{i,s},P_{i}^{*}, Z^{r*}_{\phi,s}\right)$.
  \STATE $\!\!\!\!\!\!\!\!\!$ Update $\tilde{x}_{i,s}^{r(\kappa+1)}\!\!=\! \tilde{x}_{i,s}^{r*}$, $\forall i\in \mathcal {I}, r\in \mathcal {R}, s\!\in\! \mathcal {S}_r$.
   \STATE $\!\!\!\!\!\!\!\!\!$ Update point iteration count $\kappa\! =\! \kappa{+}1$.
\ENDWHILE
\end{algorithmic}
\end{algorithm}

{We clarify that in Step 3 of Algorithm 1, $\Psi^{(\kappa)}\left(\tilde{x}^{r*}_{i,s},P_{i}^{*}, Z^{r*}_{\phi,s}\right)$ is optimized for the given $\tilde{x}^{r(\kappa)}_{i,s}$. This yields
$\Psi^{(\kappa+1)}\!\!\left(\tilde{x}^{r(\kappa+1)}_{i,s}\!\!,P_{i}^{(\kappa+1)}\!\!, Z^{r(\kappa+1)}_{\phi,s}\right)\!\!\leqslant \!\! \Psi^{(\kappa)}\!\!\left(\tilde{x}^{r(\kappa)}_{i,s}\!\!,P_{i}^{(\kappa)}\!\!, Z^{r(\kappa)}_{\phi,s}\right)$~\cite{2018_BinaryTransformation_forconvergence,1978_Optimization_Mark}.
Thus, Algorithm 1 produces a monotone sequence of improved feasible solutions,  $\Big\{\Psi^{(\kappa)}\left(\tilde{x}^{r(\kappa)}_{i,s}\!\!,P_{i}^{(\kappa)}\!\!, Z^{r(\kappa)}_{\phi,s}\right)\Big\}$,  to the problem $\mathbf {\hat{P}^{o}}$ once $\tilde{x}^{r(\kappa)}_{i,s}$ is initialized from a feasible point, e.g., $x^{r (\kappa)}_{i,s}{=}0.5, \, \forall i\in \mathcal {I}, r\in \mathcal {R}, s\!\in\! \mathcal {S}_r$.
Moreover, $\Big\{\Psi^{(\kappa)}\left(\tilde{x}^{r(\kappa)}_{i,s},P_{i}^{(\kappa)}, Z^{r(\kappa)}_{\phi,s}\right)\Big\}$ is bounded by constraints \eqref{OptProbOrg-D} and  \eqref{OptProb2Sol2-A}. Based on these, it can be concluded that the convergence of Algorithm 1 is guaranteed, i.e.,
$\Big(\Psi^{(\kappa+1)}\!\!\left(\tilde{x}^{r(\kappa+1)}_{i,s}\!\!,P_{i}^{(\kappa+1)}\!\!, Z^{r(\kappa+1)}_{\phi,s}\right)- \Psi^{(\kappa)}\!\!\left(\tilde{x}^{r(\kappa)}_{i,s}\!\!,P_{i}^{(\kappa)}\!\!, Z^{r(\kappa)}_{\phi,s}\right)\Big)\longrightarrow~0$ as $\kappa\longrightarrow\infty$~\cite{2018_BinaryTransformation_forconvergence,1978_Optimization_Mark}.}

{
By checking the operations of Algorithm 1, we find that the same steps are executed in each iteration; thus, we focus only  on the complexity of one iteration. The dominating computational complexity of the algorithm arises from Step 3, where $\mathbf {\hat{P}^{o}}$ is solved to obtain $ P_{i}^{*}, \tilde{x}_{i,s}^{r*} ,Z_{\phi,s}^{r*} $ and $\Psi^{(\kappa)}\left(\tilde{x}^{r*}_{i,s},P_{i}^{*}, Z^{r*}_{\phi,s}\right)$. Thus, solving each iteration of Algorithm 1 requires a complexity of $\mathcal{O} \Big( \big(|\mathcal {I}|^2|\mathcal {R}|{+}2|\mathcal {I}||\mathcal {R}|{+}|\mathcal {I}|\big)^3\big(|\mathcal {I}|^2|\mathcal {R}|{+}3|\mathcal {I}||\mathcal {R}|{+}3|\mathcal {I}|{+}|\mathcal {R}|{+}1\big) \Big)$~\cite{2014_PenaltyFactor2,akram2021TCOM}.
{We find from \cite{akram2021TCOM,HBM2} that solving the algorithm for spectrum allocation with ESB requires a complexity of approximately $\mathcal{O} \Big(|\mathcal {I}|^8\Big)$. Based on this and the fact that $\mathcal{C}_{\mathrm{Alg}}\approx\mathcal{O} \Big(|\mathcal {I}|^8|\mathcal {R}|^4\Big)$ when $|\mathcal {I}|\gg|\mathcal {R}|$, we conclude that the proposed algorithm for spectrum allocation with ASB is $|\mathcal {R}|^4$ times more complex than that for spectrum allocation with ESB.
}

\section{Numerical Results}

In this section, we present numerical results to examine the performance of the proposed multi-band-based spectrum allocation scheme.
Specifically, we consider an indoor environment of size $20~\textrm{m}\times 20~\textrm{m}$, where an AP at the center of the indoor environment serves 30 users. We also consider that the spectrum that exists between $0.325~\textrm{THz}$ and $0.448~\textrm{THz}$ is used to serve the users.
We clarify that the considered spectrum spans across two TWs and each of the TW is composed of a PACSR and an NACSR. We denote the two TWs and the four regions by $\{\mathrm{TW}_{2}, \mathrm{TW}_{3}\}$ and $\{p_2,n_2,p_3\,n_3\}$, respectively, as depicted in Fig. \ref{Fig:THzBand1}.
Unless otherwise stated, the values of the other parameters used for numerical results are as follows: $B_{\textrm{g}}{=}1~\textrm{GHz}$, $B_{\textrm{max}}{=}4.5~\textrm{GHz}$,  $P_{\textrm{tot}}{=}{-}12.5~\textrm{dBm}$, $P^{\textrm{max}}{=}\frac{4P_{\textrm{tot}}}{3|\mathcal{I}|}$, $R_{\textrm{thr}}{=}2~\textrm{Gbps}$, $h_{\epsilon}{=}2~\textrm{m}$, $g_{\textrm{A}}{=}35~\textrm{dBi}$, {$g_{\textrm{U}}{=}20~\textrm{dBi}$}, $N_{0}{=}{-}174~\textrm{dBm/Hz}$, $\xi^r_{\phi,s}{=}10^{9.7},~\omega^r_{\phi,s}{=}10^{10.7}$, and~$\varsigma^r_{\phi,s}{=}10^{-3}$.

In Fig. \ref{Fig:FigB}, we plot the performance gain achieved by (i) considering ASB when the to-be-allocated spectrum of interest is composed of multiple TWs, i.e., $0\leqslant B_{s}^{r}\leqslant B_{\textrm{max}},~\forall s\in \mathcal {S}^r, r \in \mathcal{R}$, and (ii)  optimally determining the unused spectra that exists at the edge of TWs, i.e., $B^r_{\Delta}\geq0,~\forall  r \in \mathcal{R}$.
In~Fig. \ref{Fig:FigB}(a), we first observe the aggregated multiuser data rate, $R_{\textrm{AG}}=\sum_{i\in \mathcal {I}}R_{i}$, achieved by the proposed scheme with ASB while considering the state-of-the-art ESB-based spectrum allocation scheme in~\cite{akram2021TCOM} as the benchmark. {We clarify that in the benchmark scheme, (i) ESB is considered, (ii) the sub-bands are optimally assigned to users while satisfying the constraints \eqref{OptProbOrg-I1}, \eqref{OptProbOrg-I2}, and~\eqref{OptProbOrg-J}, and (iii) the transmit power is allowed to vary while satisfying the constraints~\eqref{OptProbOrg-B} and~\eqref{OptProbOrg-B2}.}
We clarify that for the two schemes, $B^r_{\Delta}$ is fixed at $B^{r}_{\ast}$, where $B^{r}_{\ast}$ denotes the bandwidth of the spectrum at the edges of each region where the molecular absorption coefficient is higher than 0.3~\cite{2020_WCNC_HangNan,akram2021TCOM}.
We observe that for all power budget values, a significantly higher $R_{\textrm{AG}}$ (between $12\%$ and $15\%$) can be achieved by adopting ASB by varying the sub-band bandwidths.
Next, we compare $R_{\textrm{AG}}$ achieved by the proposed spectrum allocation scheme with ASB when $B^r_{\Delta}\geq 0$ and $B^r_{\Delta}=B^{r}_{\ast}$, and then plot the  data rate gain of considering $B^r_{\Delta}\geq0$ relative to considering $B^r_{\Delta}=B^{r}_{\ast}$, $\Delta R_{\textrm{AG}}$. We observe that although $\Delta R_{\textrm{AG}}$ is marginal when the power budget is high, a noticeable data rate gain $\Delta R_{\textrm{AG}}$ ($\approx 5\%$) exists when the power budget constraint is stringent.
These two observations show the significance of considering $0\leqslant B_{s}^{r}\leqslant B_{\textrm{max}}$ and $B^r_{\Delta}\geq0$.
We further observe in~Fig. \ref{Fig:FigB}(b) that the average value of the unused bandwidth within regions, $E[B^r_{\Delta}]$, increases when the power budget decreases, which explains the increasing $\Delta R_{\textrm{AG}}$ when the power budget decreases.

\begin{figure}[!t]
\centering\subfloat[Aggregated multiuser throughput versus the power budget.\label{1a}]{ \includegraphics[width=0.8\columnwidth]{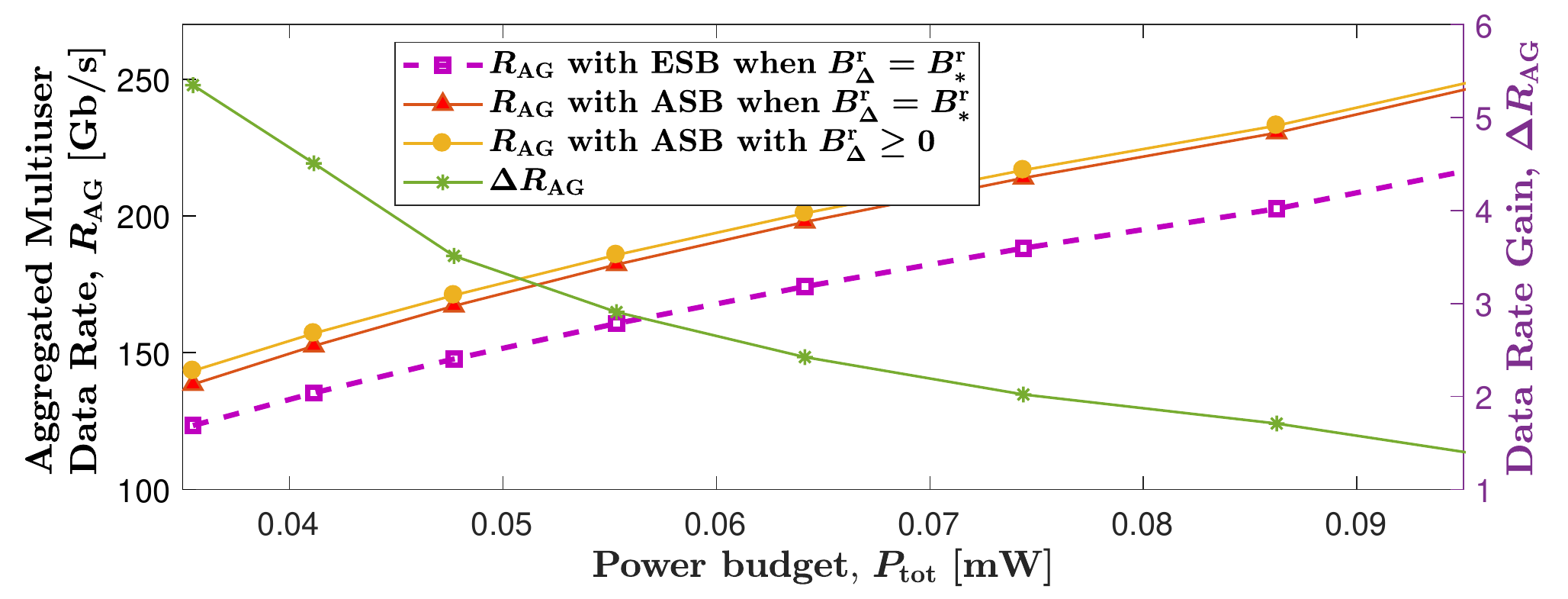}} \vspace{2mm} \hspace{10 mm}
\subfloat[{The average of unused bandwidth versus the power budget.}
\label{1b} ]{\includegraphics[width=0.8\columnwidth]{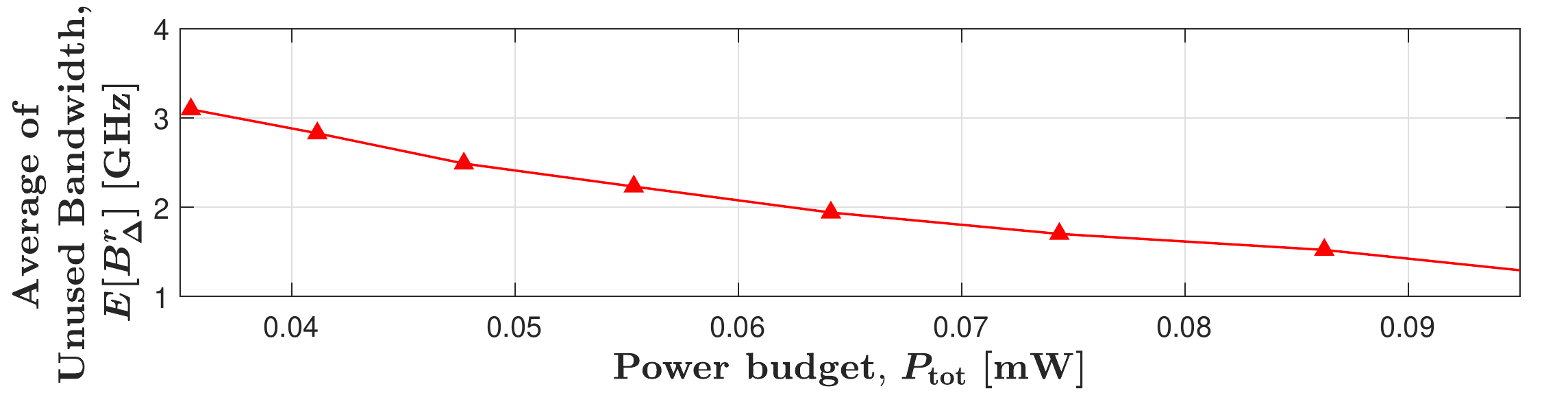}}
  \caption{Illustration of the performance achieved by the proposed spectrum allocation scheme.}\label{Fig:FigB}
\end{figure}

\begin{figure}[!t]
\centering
\includegraphics[width=0.8\columnwidth]{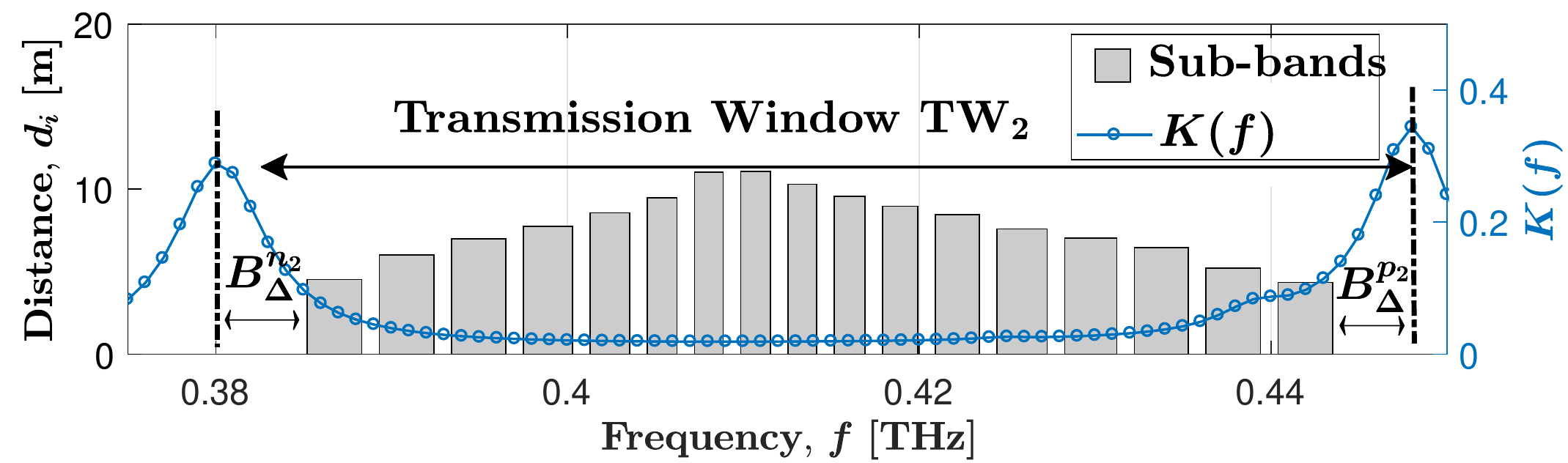}
\vspace{-4mm}
\caption{Illustration of the allocation of spectral resources within the transmission window TW2.}\label{Fig:FigC} 
\end{figure}

In Fig. \ref{Fig:FigC}, we show how the spectral resources are allocated according to the proposed multi-band-based allocation scheme. We
plot the allocation of spectral resources within the transmission window, $\mathrm{TW}_2$. We note that the gray bars represent the sub-bands and the width and the height of the bars represent the bandwidth of the sub-bands and the transmission distance of the users assigned to the sub-bands, respectively.
We first observe that, according to the optimal sub-band assignment configuration, 
the users with longer distances are assigned to center sub-bands where the absorption loss is low while the users with shorter distances are assigned to edge sub-bands where the absorption loss is high.
{This sub-band assignment configuration is similar to that considered in~\cite{HBM2,2020_Chong_InfoCom_DABM} for the scenario with ESB.}
This shows that, since the absorption loss increases exponentially with distance, the data rate gain achieved by the users with longer distances, through occupying the center sub-bands as compared to occupying the edge sub-bands, is higher than the data rate loss incurred by the users with shorter distances, through occupying the edge sub-bands as compared to occupying the center sub-bands.
We also observe that although the edge sub-bands experience higher absorption losses, a wider bandwidth is allocated to the edge sub-bands compared to the center sub-bands, since the edge sub-bands are occupied by the users with shorter distances.

In order to examine the impact of different spectrum of interest on the proposed spectrum allocation scheme,  we plot $R_{\textrm{AG}}$ versus the number of regions, $|\mathcal{R}|$, in Fig. \ref{Fig:FigA}.
Here, the regions corresponding to $|\mathcal{R}|$ of 1, 2, 3, and 4 are $\{p_2\}$, $\{p_2{+}n_2\}$, $\{p_2{+}n_2{+}p_3\}$, and $\{p_2{+}n_2{+}p_3+n_{3}\}$, respectively.
For the sake of fairness, we made the number of users in the system to be proportional to the total available bandwidth within the spectrum of interest.
More precisely, the number of users corresponding to $|\mathcal{R}|$ of 1, 2, 3, and 4 are 10, 17, 27, and 30, respectively.
We first observe that $R_{\textrm{AG}}$ achieved by our proposed scheme with ASB is equal to that achieved by the scheme in \cite{akram2021TCOM}, which is only applicable when the spectrum of interest is entirely within a region, i.e., $|\mathcal{R}|{=}1$. This shows that the scheme investigated in~\cite{akram2021TCOM} is a special case of our proposed multi-band based spectrum allocation scheme in this work. 
We also observe that the proposed spectrum allocation scheme with ASB outperforms the state-of-the-art spectrum allocation scheme with ESB by $9\%{-}13\%$ for all  values of $|\mathcal{R}|$.
Furthermore, we observe that an additional gain can be achieved by considering $B^r_{\Delta}\geq0$ as compared to $B^r_{\Delta}=B^{r}_{\ast}$.
Finally, we observe that $R_{\textrm{AG}}$ for the proposed spectrum allocation scheme improves when the upper bound on sub-band bandwidth, $B_{\textrm{max}}$, increases. This is expected since a larger $B_{\textrm{max}}$ gives more room to exploit the ASB capability, which improves $R_{\textrm{AG}}$.

\begin{figure}[!t]
\centering
\includegraphics[width=0.8\columnwidth]{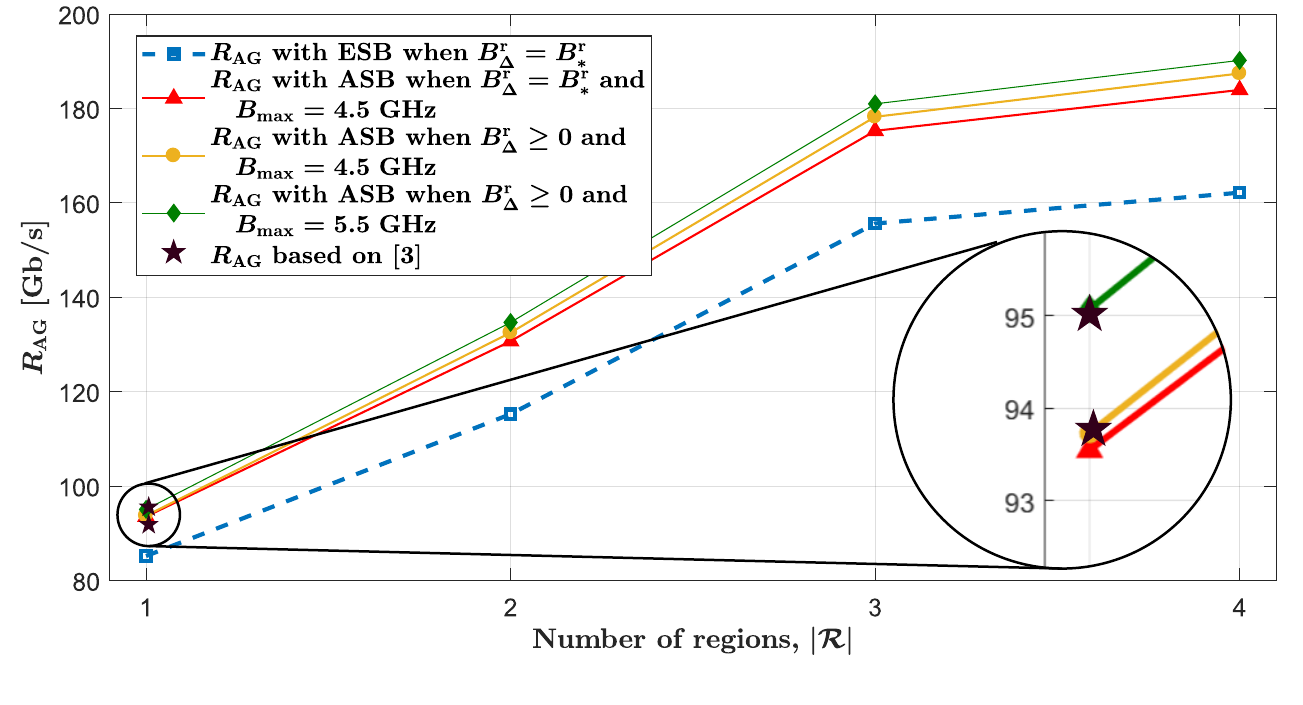}
\vspace{-6mm}
\caption{Aggregated multiuser throughput versus the number of regions within the spectrum of interest. }\label{Fig:FigA} 
\end{figure}

\begin{figure}[!t]
\centering
\includegraphics[width=0.8\columnwidth]{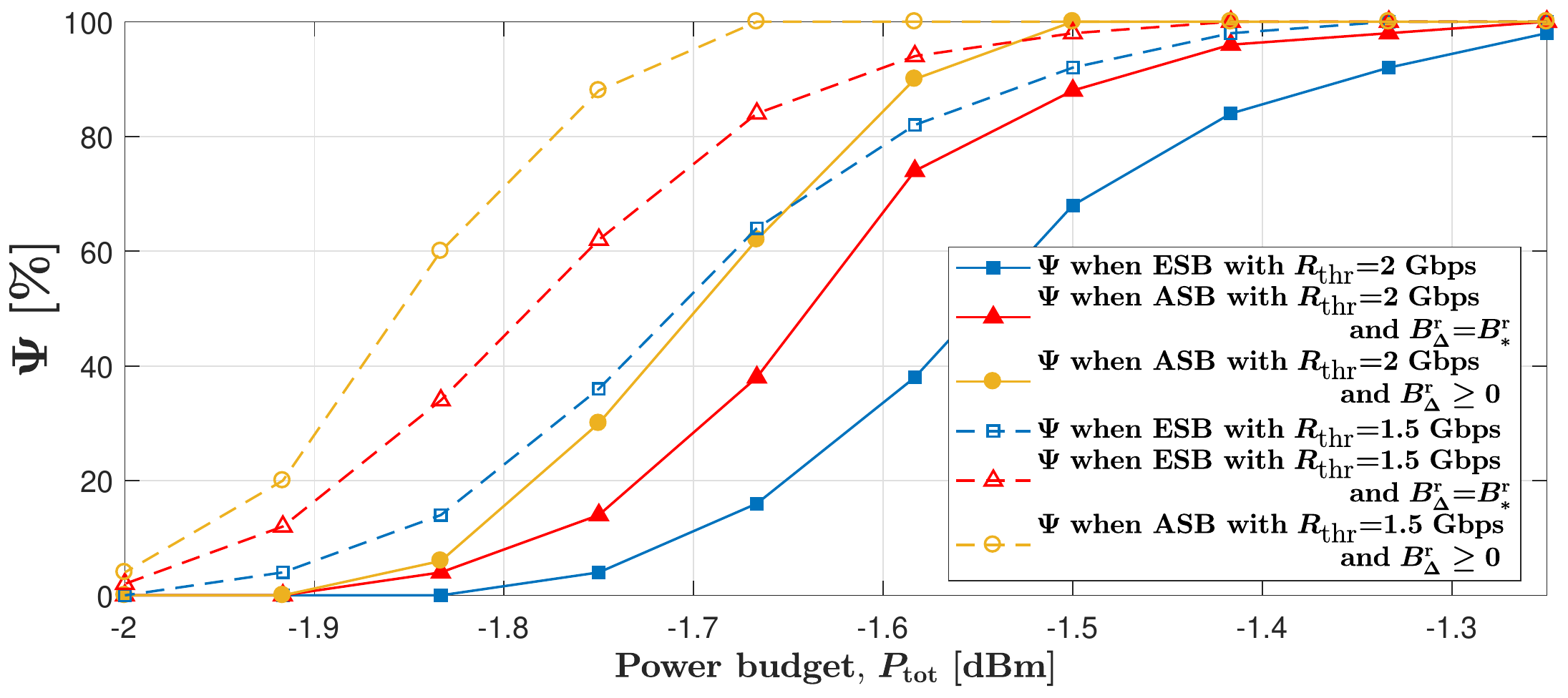}
\vspace{-4mm}
\caption{{Illustration of the variation of the percentage of the simulation trails for which the problem $\mathbf {P^{o}}$ is feasible, $\Psi$, versus the power budget.
}
}
\label{Fig:FigD} 
\end{figure}

{
Finally, we illustrate the improvement in the feasibility region of the problem  $\mathbf {P^{o}}$ that occurs as a result of considering ASB and $B^r_{\Delta}\geq0$. To this end, Fig.~6 plots the percentage of simulation trails for which the problem $\mathbf {P^{o}}$ is feasible, $\Psi$, for the benchmark scheme with ESB and the spectrum allocation scheme with ASB when $B^r_{\Delta}\geq0$ and $B^r_{\Delta}=B^{r}_{\ast}$.
We first observe that $\Psi$ increases when power budget, $P_{\textrm{tot}}$, and rate threshold, $R_{\textrm{thr}}$, increases and decreases, respectively. This is expected because the increase in $P_{\textrm{tot}}$ and the decrease in $R_{\textrm{thr}}$ will lead to the constraints~(10b) and~(10d) in $\mathbf {P^{o}}$ to be less stringent, thereby leading to more opportunities to attain a feasible solution to $\mathbf {P^{o}}$.
Second, we observe that our consideration of ASB increases $\Psi$ as compared to considering ESB.
Third, we observe that our consideration of $B^r_{\Delta}\geq0$ increases $\Psi$ as compared to considering $B^r_{\Delta}=B^{r}_{\ast}$. The second and third observations are due to the fact that 
considering ASB and $B^r_{\Delta}\geq0$ allows sub-band bandwidth to vary and some spectra that exist at the edges of TWs to be avoided from utilization during spectrum allocation, respectively, if that can yield a feasible solution point. 
This again shows the importance of our considerations of ASB and $B^r_{\Delta}\geq0$ during spectrum allocation.}

\section{Conclusion}

This paper investigated the benefits of ASB and optimally
determining the unused spectra at the edges of TWs during multi-band-based spectrum allocation in THzCom systems where the to-be-allocated spectrum of interest is composed of multiple TWs. We first formulated an optimization problem, with the main focus on spectrum allocation. Thereafter, to analytically solve the formulated problem, we proposed transformations, as well as modifications, and developed iterative algorithms based on the SCA technique. Using numerical results,  we showed that the proposed spectrum allocation scheme with ASB achieves a significantly higher data rate compared to the multi-band based spectrum allocation with ESB.  We also showed that an additional gain can be obtained by optimally determining the unused spectra at the edges of TWs as compared to  avoiding using pre-defined spectra that exist at the edges of TWs, especially when the power budget constraint is stringent. 

{
We clarify that the solution proposed in this work  is only applicable when the molecular absorption coefficient  within the TWs can be modeled using exponential functions of frequency with minimal approximation errors. Nevertheless, to harness the potentials of the huge available bandwidths of THz band, it would be beneficial to develop a solution for the spectrum allocation with ASB when the molecular absorption coefficient varies in a highly non-linear manner   within the TWs and cannot be modeled using exponential functions of frequencies, which is the case in many spectrum regions within the THz band.}

 \bibliographystyle{IEEEtran}

\end{document}